\documentclass[12pt]{iopart}
\usepackage{iopams}
\usepackage{graphicx}		
\usepackage{xcolor}

\begin{document}
\title[Simplified landscapes for optimization of shaken lattice interferometry]{Simplified landscapes for optimization of shaken lattice interferometry}

\author{C A Weidner and D Z Anderson}

\address{Department of Physics and JILA, University of Colorado, Boulder, Colorado, 80309-0440, USA}

\begin{abstract}
Motivated by recent results using shaken optical lattices to perform atom interferometry, we explore splitting of an atom cloud trapped in a phase-modulated (``shaken'') optical lattice. Using a simple analytic model we are able to show that we can obtain the simplest case of $\pm2\hbar k_\mathrm{L}$ splitting via single-frequency shaking. This is confirmed both via simulation and experiment. Furthermore, we are able to split with a relative phase $\theta$ between the two split arms of $0$ or $\pi$ depending on our shaking frequency. Addressing higher-order splitting, we determine that $\pm6\hbar k_\mathrm{L}$ splitting is sufficient to be able to accelerate the atoms in counter-propagating lattices. Finally, we show that we can use a genetic algorithm to optimize $\pm4\hbar k_\mathrm{L}$ and $\pm6\hbar k_\mathrm{L}$ splitting to within $\approx0.1\%$ by restricting our optimization to the resonance frequencies corresponding to single- and two-photon transitions between Bloch bands.
\end{abstract}
\vspace{2pc}
\noindent{\it Keywords}: Atomic physics, optical lattices, atom optics, atom interferometry
\maketitle

\section{Introduction}
\label{sec:intro}

The control of quantum mechanical systems is of interest in a variety of applications, among them quantum computing and atom interferometry. The pioneering work in \cite{Meystre2001, Meystre2001b} showed that by genetic optimization of the lattice phase modulation (or ``shaking'') function, one can precisely control the atoms' final state after shaking. We extended this idea to perform so-called ``shaken lattice interferometry'' in which the quantized momentum states of the atoms trapped in a shallow optical lattice were transformed and made to undergo a conventional interferometry sequence of splitting, propagation, reflection, reverse propagation, and recombination \cite{Anderson2017, Anderson2017b}. Optical lattices have been used for atom interferometry in Raman- or Bragg-based light-pulse schemes \cite{Pritchard2009} and to accelerate interrogated atoms using Bloch oscillations \cite{Robins2014}. In a Michelson configuration \cite{Wu2005}, one-dimensional shaken lattice interferometry was shown to have a sensitivity to applied acceleration that scales as the square of the interrogation time $T_\mathrm{I}$. Furthermore, its sensitivity can be tuned to the signal of interest (e.g. an AC acceleration signal). Atoms have been held in lattices for times on the order of tens of seconds \cite{Greiner2015}, so a shaken lattice interferometer has the possibility of achieving similar interrogation times.

In this paper we take a different approach than the usual Floquet analysis \cite{Holthaus2015, Eckardt2017, Bloch2017} to the dynamics of a shaken lattice system. We wish to explore in detail how one can shake an optical lattice to transform the wavefunction of atoms trapped in the lattice. Specifically, we seek to reduce the dimensionality of the shaking control landscape. The motivation for this is twofold: first, by simplifying the optimization landscape we can improve the efficiency of learning \cite{Rabitz2011}. This is particularly important in experiments limited by drift or finite cycle times. Second, we wish to limit heating and decoherence in the shaken lattice system. Recent work has shown that atoms in a shaken lattice are susceptible to decoherence \cite{Arimondo2007, Tino2008, Ferrari2009, Arimondo2009} when shaken at certain amplitudes $\alpha$ and frequencies $\omega$, both in the presence and absence of a signal. Furthermore, shaking of a BEC trapped in an optical lattice has been shown to cause heating in the condensate due to atom-atom interactions \cite{Bloch2017, Eckardt2017}. Atom scattering into transverse modes has also been shown to be deleterious \cite{Mueller2015, Mueller2015b}. Therefore, it is of interest to analyze the lattice shaking protocols that drive these state-to-state transitions and reduce the subspace to eliminate deleterious shaking frequencies.

The desired transformation considered in this paper is the first step of shaken-lattice-based interferometry. That is, we wish to start with atoms in the ground state of the lattice and transform them to a ``split'' state with an error less than $1\%$. The split state is defined such that the atoms equally populate two momentum states with the same magnitude but opposite sign. In particular we consider the simplest cases of splitting the atom population equally into the $\pm 2n\hbar k_\mathrm{L}$ states for $n = 1,2,$ and $3$. Here, we define the lattice wavenumber $k_\mathrm{L} = 2\pi/\lambda_\mathrm{L}$ for a lattice wavelength $\lambda_\mathrm{L}$. In general the $n$th order split state $|\psi_n(p,\theta)\rangle$ may have a relative phase $\theta$ between the two counterpropagating momentum components. That is, $|\psi_n(2n\hbar k_\mathrm{L},\theta)\rangle = e^{i\theta}|\psi_n(-2n\hbar k_\mathrm{L},\theta)\rangle$. It is important to note that one is not limited to interferometric operations such as beamsplitting when using the shaken lattice technique, but our work focuses primarily on this application.

For the simplest case of $2\hbar k_\mathrm{L}$ splitting we show that if the lattice is shaken at frequencies near the Bloch band $0$ to band $1$ transition, we can split the atom wavefunction to within the desired error. This transition gives a relative phase difference of $\pi$ between the two momentum states in the resulting split state. If we shake the lattice at half of the band $0$ to band $2$ transition frequency, we can split the atoms to within $1\%$ error with zero relative phase between the two momentum states. In each case, the simulation results are backed up by analytics. This simple shaking scheme is not suitable for higher-order splitting because the transition rate between bands drops precipitously as we transition from band $0$ to higher-lying bands. However, we find that if we optimize splitting via a genetic algorithm using only the band-to-band transition frequencies, we can achieve $4$ and $6\hbar k_\mathrm{L}$ splitting within $1\%$.

We show that after splitting to third-order ($\pm 6\hbar k_\mathrm{L}$) we can load the atoms into counter-propagating moving lattices and accelerate them, potentially achieving acceleration sensitivities that scale as $T_\mathrm{I}^3$ \cite{Robins2014}. Thus, we consider splitting only up to third order in this work.

The paper is structured as follows: In section \ref{sec:bands}, we motivate the description of the lattice dynamics in terms of the Bloch states and describe the split state in terms of these states. Section \ref{sec:pm_analytics} will describe an analytic treatment of the problem. Section \ref{sec:singlefreq_split} discusses the efficacy of $2\hbar k_\mathrm{L}$ and higher-order splitting with single-frequency shaking. Section \ref{sec:GA_bands} shows results of optimization with a genetic algorithm where we restrict ourselves to shaking at the band-to-band transition frequencies. Section \ref{sec:conc} concludes.

\section{\label{sec:bands} Bloch decomposition of the split state}

For the simulation results presented in this paper we will make the following assumptions: First, we assume that the atoms are delocalized in the lattice, i.e. in a superfluid state \cite{Bloch2002}. We will assume that the atoms are initialized with quasimomentum $q = 0$ and this quasimomentum does not change. Finally, we assume the atoms are non-interacting and that the lattice is infinite.

\begin{figure}[ht!]
\includegraphics[scale = .555]{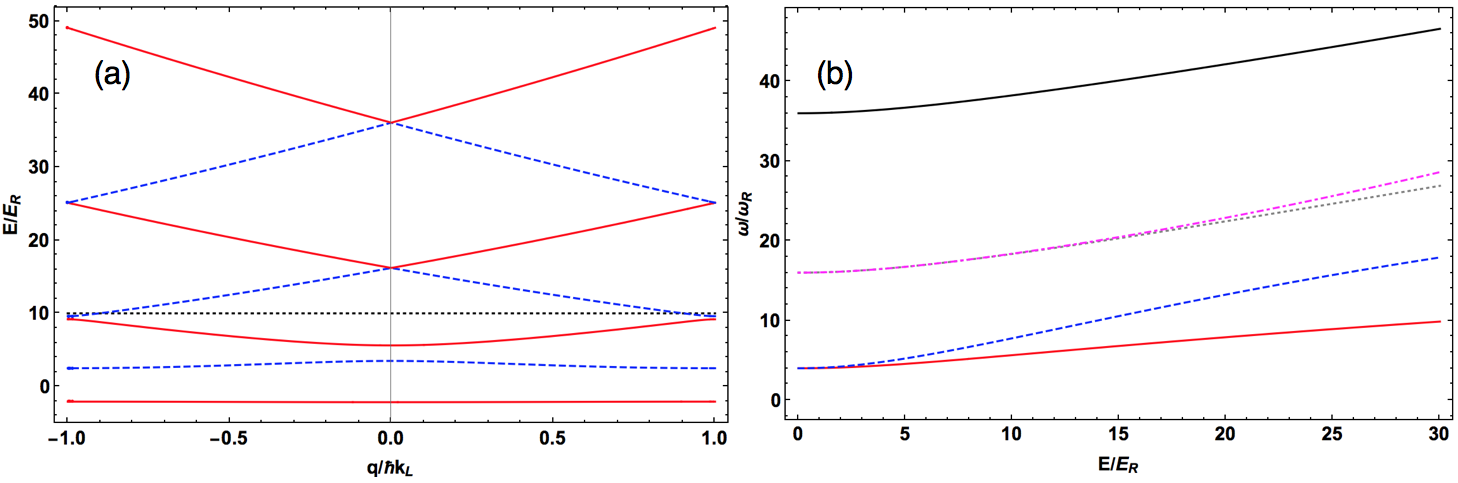}
\caption{\label{fig:bands} (a) Band energy (as a fraction of the recoil energy $E_\mathrm{R}$) versus quasimomentum $q$ (in units of $\hbar k_\mathrm{L}$) for the first seven bands, starting with $r = 0$ and ending with $r = 6$. Bands with even parity at $q = 0$ are shown with red solid lines and odd parity bands are shown with blue dashed lines. The black dotted line denotes the lattice depth. (b) Band-to-band transition frequencies (in units of $\omega_\mathrm{R} = E_\mathrm{R}/\hbar$) versus lattice depth $V_0$ (in units of $E_\mathrm{R}$) for the transition from band $r = 0$ to $r' = 1$ (red, solid), $2$ (blue, dashed), $3$ (gray, dotted), $4$ (magenta, dotted), and $5$ (black). The band $6$ transitions are almost degenerate with the band $5$ transitions for the entire of depths considered here, so the $r' = 6$ curve would completely overlap the $r'=5$ curve.}
\end{figure}

Because we are working in the regime where the atoms are delocalized in an infinite lattice, the Bloch states $|\Psi^{(q)}_r\rangle$ are a convenient basis for calculations, where $r$ denotes the band number and $q$ is the quasimomentum. For the simulations done in this paper the lattice depth was chosen to be $V_0 = 10 E_\mathrm{R}$, where the recoil energy $E_\mathrm{R} = \hbar^2k_\mathrm{L}^2/2m$ for atoms with mass $m$. The band energy $E$ is plotted against the quasimomentum $q$ in figure \ref{fig:bands}(a). The atoms begin in the state corresponding to the ground Bloch band $r = 0$ with $q = 0$. Since we assume that the quasimomentum is always zero we will suppress the index $q$ in what follows and denote the Bloch states $|\Psi^{(0)}_r\rangle$ as simply $|r\rangle$.

The Bloch states at zero quasimomentum populate only the $2n\hbar k_\mathrm{L}$ momentum states (for integer $n$). We expect then that in our model lattice modulation does not transfer momentum out of these states \cite{Meystre2001}. This is confirmed by simulation \cite{Anderson2017}. Therefore a complete description of the atom wavefunction can be given by the amplitudes and (relative) phases of the wavefunction in these quantized momentum states. Of particular interest is the relative phase $\theta$ between the two momentum components of the split state, as defined in section \ref{sec:intro}.

Experimentally one cannot determine these relative phases from time-of-flight images, as we only have access to the probability amplitudes in these experimental measurements. As such, we define a vector $\vec{P}$ with components $P_n$ containing the probability of finding an atom in the $2n\hbar k_\mathrm{L}$ momentum state \cite{Anderson2017}. If we consider an ensemble of atoms, this vector gives the relative population of atoms in each momentum state. In practice because higher-order momentum states are negligibly populated, we can truncate $|n|$ at $N = 5$. We can then define an ``error'' $E_\mathrm{ab}$ between two momentum states described by vectors $\vec{P}_\mathrm{a}$ and $\vec{P}_\mathrm{b}$ as 
\begin{equation}
\label{eq:err}
E_\mathrm{ab} = \bigg (1-\frac{\vec{P}_\mathrm{a}\cdot\vec{P}_\mathrm{b}}{|\vec{P}_\mathrm{a}||\vec{P}_\mathrm{b}|} \bigg )\times 100\%.
\end{equation}
From equation \eref{eq:err}, we see that the more similar two states are, the lower the error $E_\mathrm{ab}$. Note that if we are comparing any state to the split state, $E$ will be independent of $\theta$ and is thus a more useful quantity to look at when considering the results in the context of what is experimentally observable. Thus, we use this expression as an error measure to quantify how well our optimization algorithm is doing.

\begin{figure}[ht!]
\includegraphics[scale = .32]{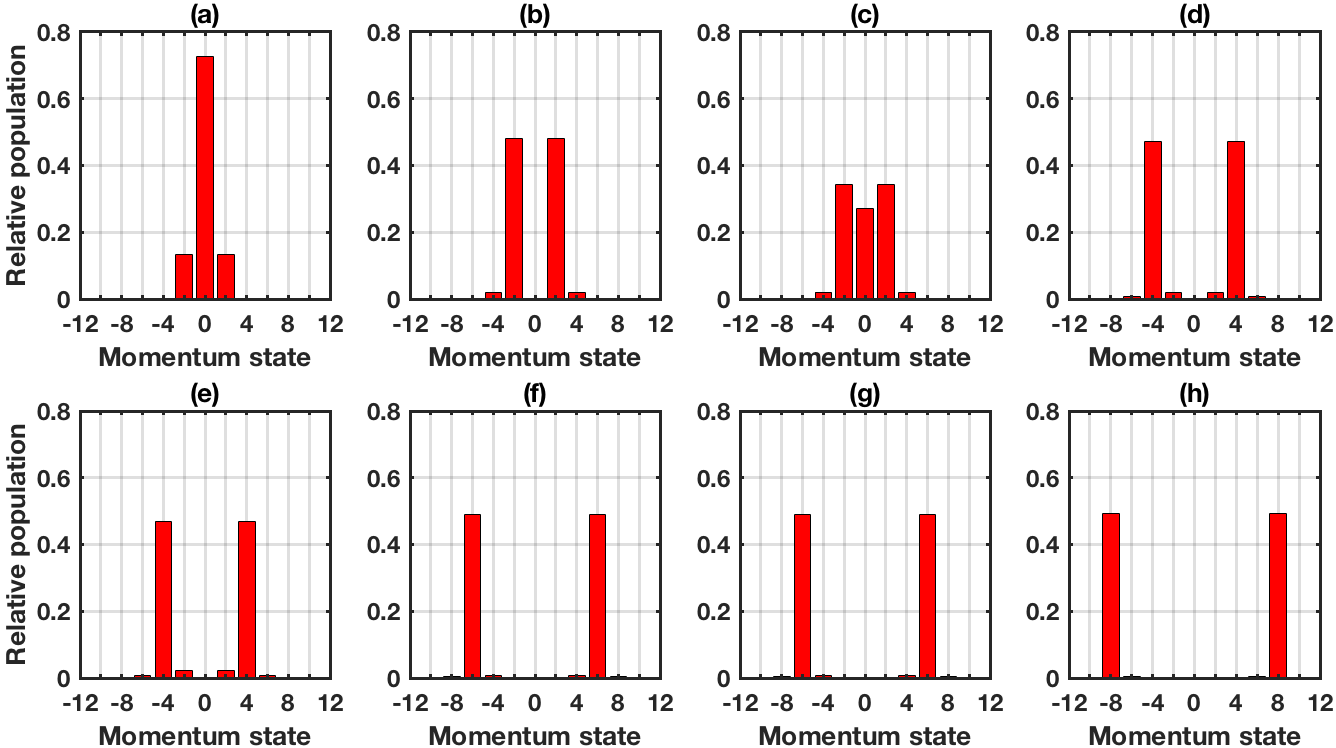}
\caption{\label{fig:bands_mom_pop} (a-h, in order of increasing band number from $r = 0$ to $7$) Momentum state populations and energies for the first 8 Bloch states $|r\rangle$ for atoms trapped in a lattice with $V_0 = 10E_\mathrm{R}$. Note that for $r > 0$ the states begin to resemble split states of higher and higher orders $n$.}
\end{figure}

\begin{figure}[ht!]
\includegraphics[scale = 0.54]{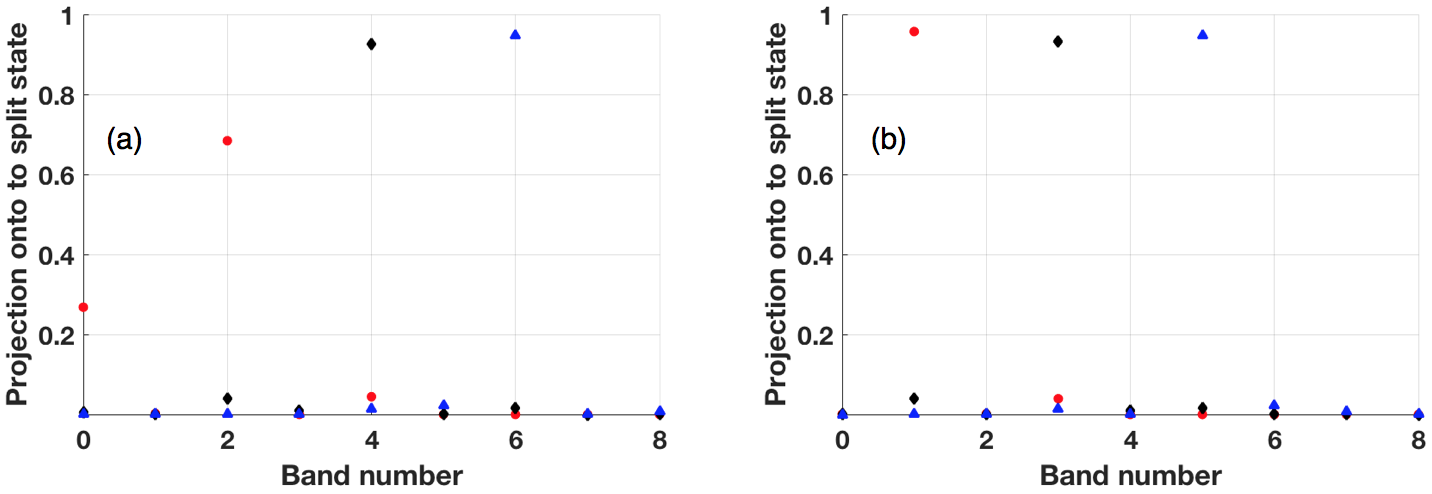}
\caption{\label{fig:band_ip} Value of the integral in equation \eref{eq:inner_prod} versus the state $|r\rangle$ corresponding to the band $r$ for splitting order $n = 1$ (red circles), $n = 2$ (black diamonds), and $n = 3$ (blue squares) for (a) $\theta = 0$ and (b) $\theta = \pi$. As the band number increases the Bloch wavefunctions look more and more like the split states, with the relative phase $\theta$ between the $\pm 2n\hbar k_\mathrm{L}$ momentum states equal to $\theta = 0$ ($\pi$) for even (odd) band numbers. Thus, alternating bands have relative phases $\theta$ of $0$ and $\pi$, depending on the band parity as shown in figure \ref{fig:bands}(a).}
\end{figure}

For bands $r>0$ there is considerable similarity between the Bloch states and split states of various orders. This is most easily seen when one looks at the momentum state population of the Bloch states corresponding to different bands, as shown in figure \ref{fig:bands_mom_pop} \cite{Phillips2002}. Interestingly, there are two separate Bloch states at different band energies that roughly correspond to each split state. To glean further insight, we calculate the inner product $D_{nr}$ between the $n$th order split state $\psi_n(p,\theta)$ and the the state $|r\rangle$ as
\begin{equation}
\label{eq:inner_prod}
D_{nr} = |\langle r|\psi_n(p,\theta)\rangle|^2.
\end{equation}
From figure \ref{fig:band_ip}, we see that the difference between two Bloch states corresponding to bands $l>0$ and $l' = l + 1$ is that the lower-energy state $|l\rangle$ has a relative phase difference $\theta = \pi$ between the $\pm (l+1)\hbar k_\mathrm{L}$ states, and the higher-energy band $|l'\rangle$ is almost identical, except $\theta = 0$ (and thus the two states are orthogonal). This is commensurate with the fact that states corresponding to adjacent bands have opposite parity.

In the rest of this paper we will be referring to state-to-state transitions between different bands. The transition frequencies for transitions from the ground band to the first $5$ excited bands are shown in figure \ref{fig:bands}(b). For an example of the energy scales at play here, we tabulate the transition frequencies $f_{r,r'}$ between two bands $r$ and $r'$ in Table \ref{tab:bandtrans}. The frequencies given in Table \ref{tab:bandtrans} assume that we are working with $^{87}\mathrm{Rb}$ atoms (as in Sec. \ref{subsec:expt}) at a lattice depth of $V_0 = 10E_\mathrm{R}$. We see that the band transitions lie between $0$ and $121$~kHz, and this is the regime in which our driving is simulated.

\begin{table}[ht!]
\caption{Band transition frequencies $f_{r,r'}$, $V_0 = 10 E_\mathrm{R}$. The single (double) asterisk marks frequencies with matrix element overlaps $M^{(c)} (M^{(s)})>0.1$ (see figure \ref{fig:mat_el_total} and section \ref{sec:pm_analytics}).} 
\renewcommand{\arraystretch}{1.2}
\centering 
\begin{indented}
\item[]\begin{tabular}{c c c} 
\br
~~Band $n$~~ & ~~Band $m$~~ &~$f_{r,r'}$ (kHz)~  \\ 
\mr
0 & 1 & 17.89** \\
0 & 2 & 24.61* \\
0 & 3 & 58.14 \\
0 & 4 & 58.25 \\
0 & 5 & 121.19 \\
1 & 2 & 6.72** \\
1 & 3 & 40.25* \\
1 & 4 & 40.36** \\
1 & 5 & 103.30 \\
2 & 3 & 33.53** \\
2 & 4 & 33.64* \\
2 & 5 & 96.58 \\
3 & 4 & 0.10 \\
3 & 5 & 63.0* \\
4 & 5 & 62.9** \\
\br
\end{tabular}
\end{indented}
\label{tab:bandtrans}
\end{table}

The next section will analytically explore the dynamics of atoms trapped in a shaken optical lattice. This case is somewhat difficult to solve analytically, but some basic results can be applied in the simple case of single-frequency shaking at low amplitudes. We verify these results numerically, showing that we can split the atom wavefunction with a phase difference of $0$ or $\pi$, depending on our driving frequency. Experimental results verify the numerics. From this we gain some insight on how to restrict our optimization parameters and show the results of rapid optimization of higher-order splitting in section \ref{sec:GA_bands}.

\section{\label{sec:pm_analytics} Analytics of phase modulation of an optical lattice}

The Hamiltonian for a phase-modulated (shaken) lattice with general shaking function $\phi(t)$ is written
\begin{equation}
\label{eq:H_PM}
H = \frac{p^2}{2m} + \frac{V_0}{2}\cos{\big [2k_\mathrm{L}x + \phi(t)\big ]}.
\end{equation}
For the specific case where $\phi(t) = \alpha \sin{(\omega t)}$, we can apply the Jacobi-Anger expansion to equation \eref{eq:H_PM}. Using this we can write the potential term $V(x,t)$ in equation \eref{eq:H_PM} as
\begin{eqnarray}
\label{eq:V_PM_JA}
V(x,t) = & V_0\bigg \{\cos{(2k_\mathrm{L}x)}\big [J_0(\alpha)/2 + \sum_{k = 1}^\infty J_{2k}(\alpha) \cos{(2k\omega t)} \big ] \nonumber \\&- \sin(2k_\mathrm{L}x)\sum_{k=1}^\infty J_{2k-1}(\alpha) \sin{\big [ (2k- 1)\omega t\big]} \bigg \}.
\end{eqnarray}
Equation \eref{eq:V_PM_JA} shows that we can decompose the phase modulation to a term representing the carrier (first term) and a set of amplitude modulation terms containing both sine and cosine terms. The strength of these amplitude modulation terms are given by the Bessel functions $J_k(\alpha)$ where $\alpha$ is the amplitude of the phase modulation. Because the sine terms are odd, they will drive transitions between opposite parity states while the even cosine terms will drive transitions between states with the same parity \cite{Phillips2002}.

By taking the $J_0$ term in equation \eref{eq:V_PM_JA} as the bare Hamiltonian $H_0(x)$, we can write the rest of the terms as a perturbation $H'(x,t)$. Using Fermi's golden rule, we can then write down the transition rate $\Gamma_{r,r'}$ from state $|r\rangle$ to $|r'\rangle$ resulting from shaking at a frequency $\omega$ as
\begin{eqnarray}
\Gamma_{r,r'} &= \frac{2\pi}{\hbar}V_0^2 \sum_{k=1} \bigg [J_{2k}^2(\alpha)|\langle r'| \cos{(2k_\mathrm{L} x)}| r \rangle |^2\times\delta(E_{r,r'} - 2k\hbar \omega) \nonumber \\ &+ J_{2k-1}^2(\alpha)|\langle r'| \sin{(2k_\mathrm{L} x)}| r \rangle |^2 \times\delta(E_{r,r'} - (2k-1)\hbar \omega) \bigg ] \label{eq:FGR}
\end{eqnarray}
where $E_{r,r'} = \hbar \omega_{r,r'} = E_r - E_{r'}$ is the energy difference between states $|r\rangle$ and $|r'\rangle$. The transition rate $\Gamma_{r,r'}$ is governed by the squares of the Bessel functions $J^2_k(\alpha)$ (where $\alpha$ is the amplitude of the phase modulation) and magnitudes of the transition matrix elements $|M^\mathrm{(s)}_{r,r'}|^2 = |\langle r' |\sin{(2k_\mathrm{L}x}| r\rangle|^2$ and $|M^\mathrm{(c)}_{r,r'}|^2 = |\langle r' |\cos{(2k_\mathrm{L}x} | r\rangle|^2$.

\section{\label{sec:singlefreq_split} Single-frequency splitting}

This section consists of three parts. The first subsection will show first-order single-frequency shaking results via simulation. Next, we will show experimental data that supports the simulation results. Finally, we will discuss how far we need to split to implement an accelerating lattice scheme.

\subsection{Theory and simulation of single-frequency splitting}

We can use the theory derived in the previous section to make some predictions about the effects of single-frequency shaking. As stated in Sec. \ref{sec:pm_analytics} the matrix elements $|M^\mathrm{(c)}_{r,r'}|^2$ connect states with the same parity, and the matrix elements $|M^\mathrm{(s)}_{r,r'}|^2$ connect states of opposite parity. 

For a given value of $\alpha$ the amplitude of the Bessel functions $J_k(\alpha)$ dies off as $k$ increases. For $\alpha \leq 0.3$ we can keep two terms, simplifying the potential in equation \eref{eq:V_PM_JA} to
\begin{eqnarray}
V(x,t) &= V_0\bigg [J_0(\alpha)\cos{(2k_\mathrm{L}x)}\big/2 \nonumber\\ &- J_1(\alpha)\sin{(\omega t)}\sin{(2k_\mathrm{L}x)} + J_2(\alpha)\cos(2\omega t)\cos{(2 k_\mathrm{L} x)}\bigg ]. \label{eq:V_PM_JA_simpl}
\end{eqnarray}

As with equation \eref{eq:V_PM_JA_simpl} if we keep two terms in equation \eref{eq:FGR}, we obtain

\begin{eqnarray}
\Gamma_{r,r'} &= \frac{2\pi}{\hbar}V_0^2 \bigg [J_{2}^2(\alpha)|M^\mathrm{(c)}_{r,r'}|^2\delta(E_{r,r'} - 2n\hbar \omega) \nonumber\\ &+ J_{1}^2(\alpha)|M^\mathrm{(s)}_{r,r'}|^2\delta(E_{r,r'} - (2n-1)\hbar \omega) \bigg ]. \label{eq:FGR_2term}
\end{eqnarray}

From this we see that weak shaking of the lattice at $\omega_{r,r'} = 2\pi\times f_{r,r'}$ will drive transitions between Bloch states $|r\rangle$ and $|r'\rangle$ if they have opposite parity and driving at $\omega_{r,r'}/2$ will drive transitions between $|r\rangle$ and $|r'\rangle$ if they have the same parity. In general, shaking at $\omega_{r,r'}/N$ will drive transitions between $|r\rangle$ and $|r'\rangle$ with parity determined by whether $N$ is odd or even. This is in keeping with the results in \cite{Eckardt2016} for the case of the amplitude-modulated lattice (where only like-parity transitions are allowed) and the phase-modulation results in \cite{Simonet2015}. The difference in our work is that we approach this problem from a standpoint of inducing band-to-band transitions to perform atom beamsplitting for interferometry.

The above analysis shows that if we begin in the ground state $|r = 0\rangle$ and shake at $\omega = \omega_{01}$ ($\omega_{02}/2$), we will drive odd (even) parity transitions between bands $r = 0$ and $r' = 1$ ($r' = 2$). We simulate both cases using the symmetric split-step method \cite{Steiger1982} to simulate the time-dependent Schr{\"o}dinger equation (TDSE) with the Hamiltonian in equation \eref{eq:H_PM} with a single frequency $\omega$ and amplitude $\alpha = 0.3$ for $T \approx 1$~ms.

\begin{figure}[ht!]
\includegraphics[scale = .62]{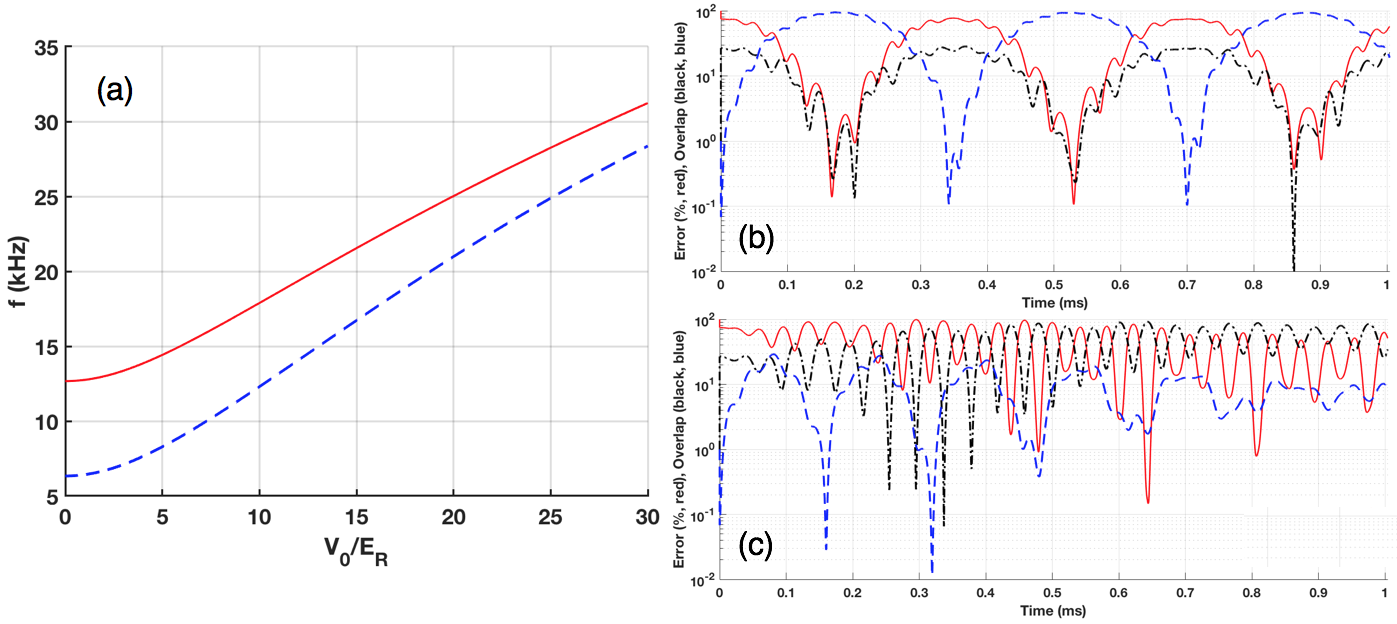}
\caption{\label{fig:shake_omega} (a) The band transition frequencies $\omega_{01}$ (red, solid) and $\omega_{02}/2$ (blue, dashed) as a function of the lattice depth. (b-c) Results of shaking simulations for a lattice depth of $V_0 = 10E_\mathrm{R}$. The percent error relative to the split state (red, see equation \eref{eq:err}), projection of the current state $|\psi\rangle$ onto the split state with $\theta = 0$ (blue, dashed, see equation \eref{eq:inner_prod}) and $\theta = \pi$ (black, dot-dashed), plotted versus shaking time for a shaking frequency of (b) $\omega_{01}$ and (c) $\omega_{02}/2$ and a shaking amplitude of $\alpha = 0.3$~rad.}
\end{figure}

The band transition frequencies $\omega_{01}$ and $\omega_{02}/2$ are plotted in figure \ref{fig:shake_omega}(a) versus the lattice depth. Given $V_0 = 10E_\mathrm{R}$, results for odd parity shaking at $\omega = \omega_{01} = 2\pi\times 17.88$~kHz are shown in figure \ref{fig:shake_omega}(b), and results for even parity shaking at $\omega_{02}/2 = 2\pi\times 12.3$~kHz are shown in \ref{fig:shake_omega}(c). For the simulation results presented in figure \ref{fig:shake_omega}(b-c), at each timestep we calculate the percent error relative to the split state as in equation \eref{eq:err} and the inner product between the simulated state $|\Phi(t)\rangle$ at time $t$ and the first order split state $|\psi_1(p,\theta)\rangle$ as in equation \eref{eq:inner_prod}.

We see that when the percent error is lowest in figure \ref{fig:shake_omega}(b), the projection of the state $|\psi\rangle$ onto the split states is highest for the split state $|\psi_1(p, \theta = \pi)\rangle$. This shows that we are in fact splitting with relative phase $\theta = \pi$ between the two split arms. Conversely, in figure \ref{fig:shake_omega}(c) we achieve splitting with $\theta = 0$. Thus, by controlling the shaking frequency we can control the relative phase between the two split arms.

\begin{figure}[ht!]
\includegraphics[scale = 0.6]{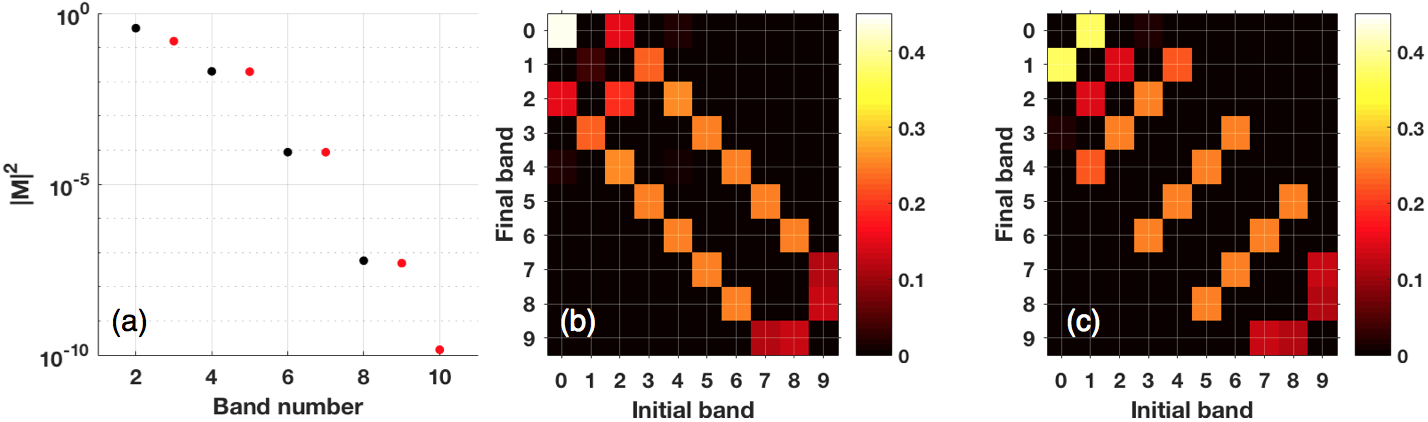}
\caption{\label{fig:mat_el_total} (a) The matrix elements $|M^\mathrm{(c)}_{r,r'}|^2$ (red) and $|M^\mathrm{(s)}_{r,r'}|^2$ (black) for band-to-band transitions plotted versus final band $r'$. (b-c) Plots of the matrix elements (b) $|M^\mathrm{(c)}_{r,r'}|^2$ and (c) $|M^\mathrm{(s)}_{r,r'}|^2$ for band-to-band transitions from bands $r$ (x-axis) to $r'$ (y-axis). The colorbar on the right gives the magnitude of the transition matrix element.}
\end{figure}

For higher amplitudes first-order perturbation theory becomes less and less applicable, and we can no longer use Fermi's Golden rule to accurately describe the physics of the situation. In this case we must keep more terms in the Jacobi-Anger sums of equation (\ref{eq:V_PM_JA_simpl}) and go to higher orders in perturbation theory. Furthermore, we cannot use this simple picture to obtain higher-order splitting. This is due to the fact that the matrix elements $|M^\mathrm{(c)}_{r,r'}|^2$ and $|M^\mathrm{(s)}_{r,r'}|^2$ become relatively small as we consider transitions from the state $|r = 0\rangle$ to higher-lying states with $|r>2\rangle$. This is shown in figure \ref{fig:mat_el_total}(a) where for higher-band transitions the relevant matrix element is at least one order of magnitude below the lower-band transitions. As such, the transition strength is much lower and the transitions become less favorable.

However, we can make transitions from the ground state $|r = 0\rangle$ to an intermediate state in band $r' = 1$ or $2$ and then to higher-lying states in bands $r''>2$. If we make these intermediate state transitions, the matrix elements become more favorable. This is shown in figure \ref{fig:mat_el_total}(b-c). As expected, the cosine transition matrix elements are strongest when making transitions between states in next-to-adjacent bands, but the sine matrix elements are strongest when making transitions between states in adjacent bands. Interestingly, when considering the sine matrix elements we see that it is also favorable to make transitions between states in bands $r = 1$ to $r' = 0, 2$ or $4$. We also observe that transitions between bands $r = 3$ and $r' = 2$ and $6$ are favorable, but transitions between bands $r = 3$ and $r' = 4$ are not. This is possibly due to the avoided crossing between bands $3$ and $4$ at $q = 0$ (see figure \ref{fig:bands}). We will find that the strongest transitions in figure \ref{fig:mat_el_total} are most influential can be used to simplify the optimization landscape for higher-order splitting in section \ref{sec:GA_bands}.

\subsection{Experimental results}
\label{subsec:expt}

In this section we demonstrate that the splitting schemes described in the previous subsection and shown in figure \ref{fig:shake_omega} are viable experimentally. The experimental scheme is similar to the shaken lattice inteferometry experiment described in \cite{Anderson2017b}. In the experiment we start with Bose-condensed $^{87}\mathrm{Rb}$ atoms loaded into the ground state of an optical lattice of (intentionally) unknown depth. The lattice is made by retro-reflecting an $852$~nm laser onto itself. By modulating the frequency of the lattice laser \cite{Sherson2014}, we shake the lattice for a time $T = 0.2$~ms with varying amplitude $A$ and frequency $f$. We use a computer-controlled arbitrary waveform generator (AWG) to generate a pure tone modulated by an envelope of the form $f(t) = \cos^2{(\pi t/T)}$, which allows for smooth turn-on and turn-off of the shaking, as in \cite{Anderson2017, Anderson2017b}. After $20$~ms time-of-flight, we take an absorption image of the atoms using a CCD camera and analyze the images to extract the atoms' momentum state.

\begin{figure}[ht!]
\includegraphics[scale = 0.615]{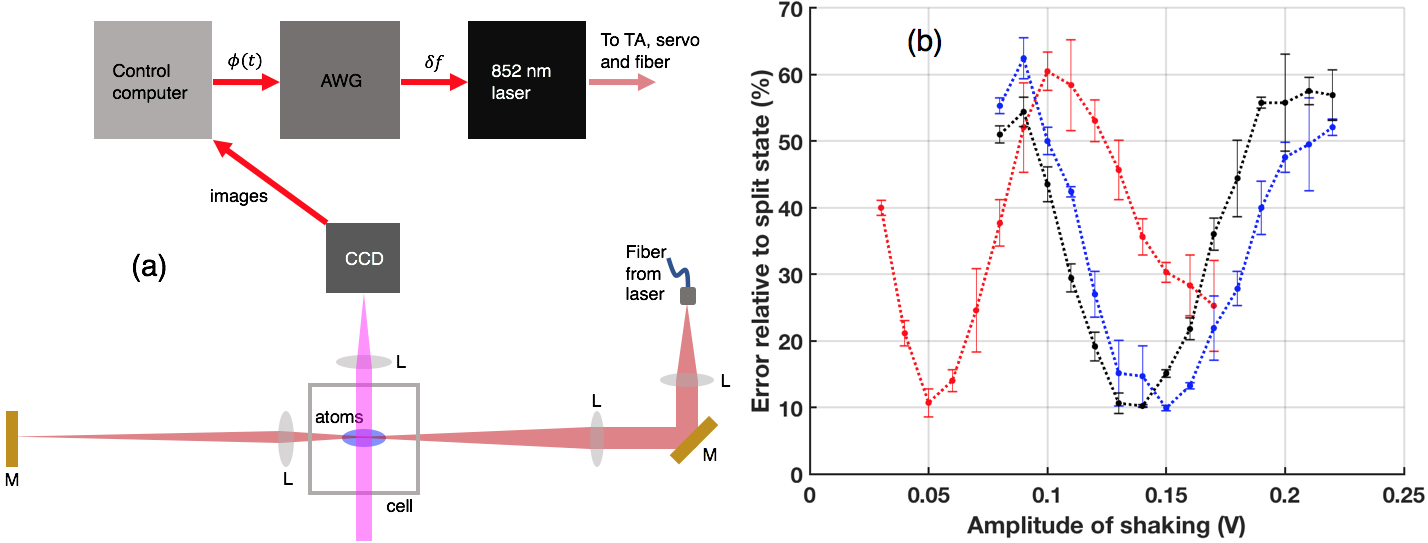}
\caption{\label{fig:expt_results} (a) A cartoon schematic of the experimental layout described in the main text. (b) Experimental results showing the percent overlap with the split state versus shaking amplitude $A$ for frequencies of $17$~kHz$\approx\omega_{02}/2$ (red), $21.5$~kHz (blue), and $22$~kHz (black), where $\omega_{01} \approx 21.7$~kHz for a lattice depth $V_0 \approx 15.3 E_\mathrm{R}$ (see figure \ref{fig:shake_omega}(a)). The points show the mean of three separate experimental runs with the same shaking function and the error bars show the standard deviation of these points.}
\end{figure}

We find that we can split the atom wavefunction to within $E\approx 10\%$ at frequencies corresponding to $\omega_{01}$ and $\omega_{02}/2$. This is shown in figure \ref{fig:expt_results}(b). We do not generally obtain splitting to better than $10\%$ due to spurious atoms detected in the $0\hbar k_\mathrm{L}$ momentum state (e.g. due to heating and imaging noise), the finite momentum spread of the condensed atoms in the lattice \cite{Esslinger2001, Raithel2009}, and the finite signal-to-noise ratio in imaging \cite{Anderson2017b}. Our experimental momentum width is about $0.6\hbar k_\mathrm{L}$, limited mostly by the tight atom-chip-based trap in which we perform our evaporation. Simulations show that for such momentum spreads we are limited to errors of about $4-8\%$, depending on our shaking frequency.

From the results shown in figure \ref{fig:expt_results}, we estimate the lattice depth to be $V_0 \approx 15.3E_\mathrm{R}$. In this case the relevant band-to-band transition frequencies are $\omega_{01} = 21.7$~kHz and $\omega_{02} = 2\times17$~kHz. This not only confirms the simulation results from the previous subsection but provides us with a reliable way to approximate the lattice depth. In our current interferometry experiments the absolute lattice depth is less important than the day-to-day and shot-to-shot stability of the lattice depth. Thus, an approximate depth combined with the stability provided by a lattice laser intensity servo is sufficient for our purposes.

\subsection{How much must we split?}

Due to the fact that the single-frequency shaking does not work as well for higher-order splitting (see figure \ref{fig:mat_el_total}), higher-order splitting is more difficult to obtain. This is the subject of section \ref{sec:GA_bands}. However, before we dive into the next section it is instructive to demonstrate how much we must split the atoms to be able to accelerate them in a moving lattice.

If we truncate the Bessel function expansion of equation \eref{eq:V_PM_JA_simpl} to two terms and do some trigonometry, we obtain
\begin{eqnarray}
V(x,t) &= 2V_0\bigg \{J_0(\alpha)\cos{(2k_\mathrm{L}x)}\big/4 + J_1(\alpha)\big [\cos{(2k_\mathrm{L}x -\omega t)} \nonumber \\ &+ \cos{(2k_\mathrm{L}x +\omega t)}\big ]\bigg \}.\label{eq:V_PM_2term}
\end{eqnarray}
Equation \eref{eq:V_PM_2term} describes a carrier lattice and two counterpropagating moving lattices with velocity $v = \pm \omega/2k_\mathrm{L}$. If we can split the atoms to a high enough order, we can trap the split atoms in one of the two moving lattices. The atoms will then move with the lattice if we accelerate and decelerate the lattice. This will allow us to obtain interferometry with $T_\mathrm{I}^3$ sensitivity to an applied signal\cite{Robins2014}. In this case the moving atoms will not be able to ``see'' the counterpropagating lattices and will thus not be affected by them\footnote{The atoms moving with one of the lattices must be in an eigenstate of the lattice, but shaking can always be modified to prepare the split atoms so that they resemble the ground state of the moving lattice with depth $V_0J_1(\alpha)$.}. In this case, the atoms in the positive (negative) momentum state will be trapped in the lattice moving with positive (negative) velocity. Then, if the lattices are accelerated by changing the shaking frequency such that the magnitude of the counterpropagating lattice velocity changes, the atoms should follow the lattices in which they are trapped. The atoms will thus accelerate as the lattices are accelerated, given that this is done slowly enough \cite{Meschede2014, Anderson2016}.

\begin{figure}[ht!]
\includegraphics[scale = .6]{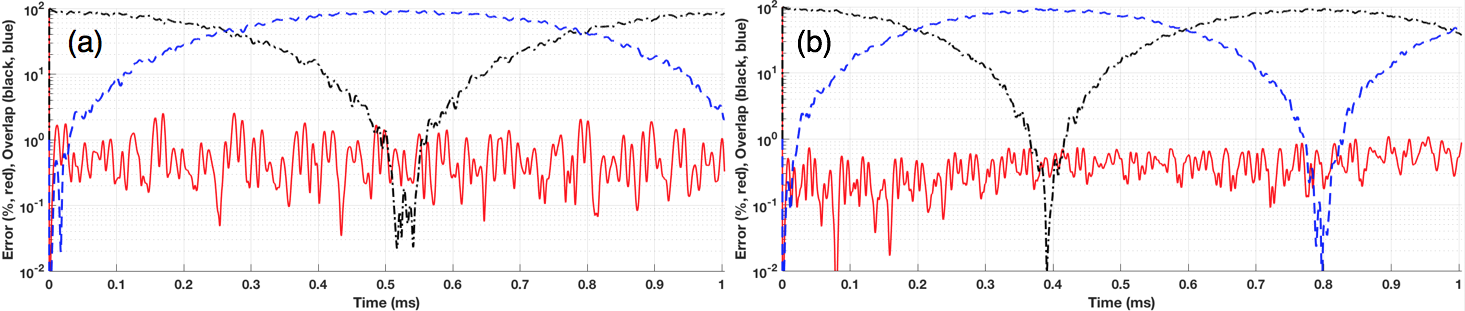}
\caption{\label{fig:accel_lattices} The percent error relative to the (a) third- ($n = 3$) and (b) fourth-order ($n = 4$) split state (red, see equation \eref{eq:err}), projection of the current state $|\Phi(t)\rangle$ onto the split state with $\theta = 0$ (blue, dashed, see equation \eref{eq:inner_prod}) and $\theta = \pi$ (black, dot-dashed), plotted versus shaking time for a shaking amplitude $\alpha = 1$~rad. In each case, there is are two counterpropagating lattices moving at velocities $v=\pm 2n\hbar k_\mathrm{L}/m$.}
\end{figure}

We find that if we begin with atoms split to third order (that is, $\pm 6\hbar k_\mathrm{L}$), we can shake the lattice at $\omega = 12\hbar k_\mathrm{L}^2/m = 12\omega_\mathrm{R}$ with $\alpha = 1$ such that the lattice is moving with $v = \pm6\hbar k_\mathrm{L}/m$ without disturbing the atom wavefunction appreciably. Here, $\omega_\mathrm{R} = E_\mathrm{R}/\hbar$ is the recoil frequency of the atoms in the lattice. In this case the atoms maintain their split state to within $\approx1\%$, as shown in figure \ref{fig:accel_lattices}(a). Furthermore, simulations show that if the atoms begin in the $\pm 8\hbar k_\mathrm{L}$ split state and are trapped in a lattice shaken at $\omega = 16\hbar k_\mathrm{L}^2/m$, the atoms will continue to maintain their splitting to within $1\%$, as shown in figure \ref{fig:accel_lattices}(b). This trend continues for even higher splitting orders.

This analysis shows that if we can achieve third-order splitting we can then accelerate the atoms in the lattice with negligible perturbation. In the next section we will show how to optimize such splitting by shaking at frequencies corresponding to band-to-band transitions.

\section{\label{sec:GA_bands} Optimization of higher-order splitting using band-to-band transitions}

This section presents results of simulations optimizing splitting protocols up to $n = 3$. Our optimization simulates the TDSE using the Hamiltonian in equation \eref{eq:H_PM} as in section \ref{sec:pm_analytics} and the optimization tries to minimize the error as given in equation \eref{eq:err}. The optimization is done via a genetic algorithm as detailed in \cite{Meystre2001, Anderson2017}. We will compare results using the full frequency bandwidth up to the $r = 0 \rightarrow r' = 5'$ transition to optimize the lattice shaking to results where only single-photon band-to-band and two-photon half-band transitions are used. By the term ``half-band transitions,'' we mean that we shake the lattice at a frequency $\omega_{r, r'}/2$ to cause even parity transitions. We know from \cite{Bloch2017} that off-resonant shaking in the presence of atom-atom interactions causes heating. Furthermore, to avoid the transverse scattering described in \cite{Mueller2015, Mueller2015b} we want to shake at single- and two-photon band-to-band resonances so that no excess energy is available for transverse scaattering. Therefore we wish to restrict our optimization to resonant transitions in order to limit the heating due to these factors. Note that as in \cite{Anderson2017} our simulations do not take interactions into account.

In the simulations presented here we shook the lattice for $T \approx 0.5$~ms and optimized for $\pm 2n\hbar k_\mathrm{L}$ splitting for $n = 1,2,$ and $3$. To ensure smooth turn-on and turn-off of the shaking function, we multiplied each shaking function by an envelope function $f(t) = \cos{(2\pi t/T)}$ \cite{Anderson2017, Anderson2017b}. Due to the inherent randomness in the GA, we ran each class of simulations $10$ separate times and took the best result for our analysis.

\begin{figure}[ht!]
\includegraphics[scale = 0.59]{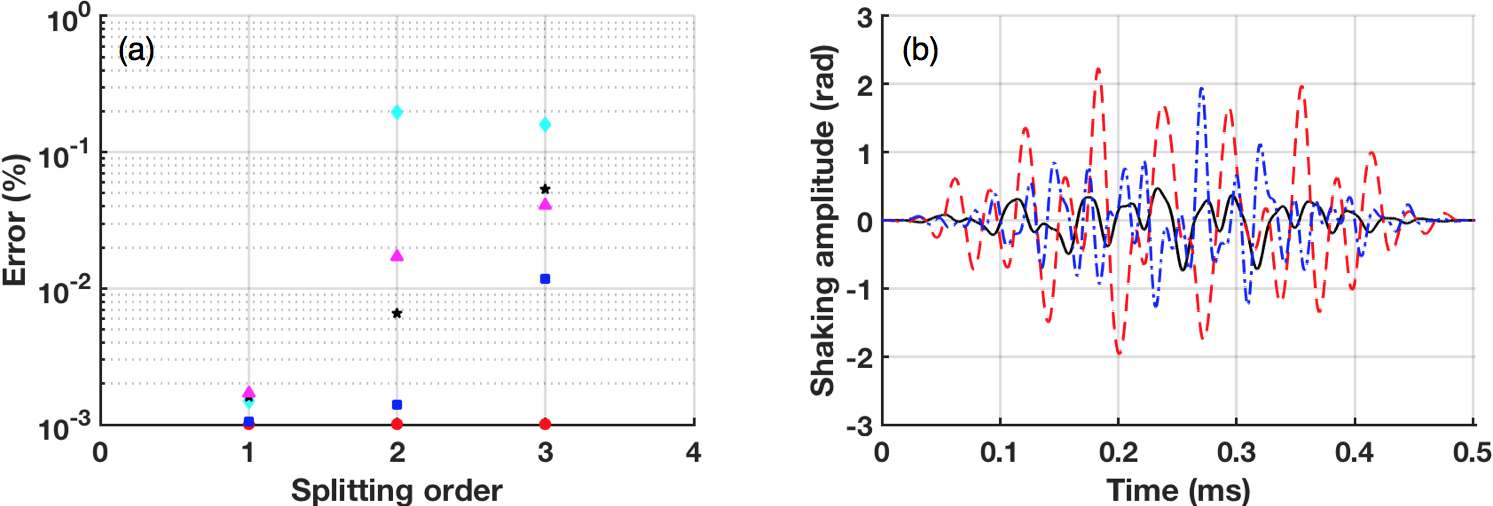}
\caption{\label{fig:GAfreqs_pct_err} (a) Percent error vs. splitting order for $n = 1,2$ and $3$ for the five simulations considerd in this work. The frequency ranges considered here are: band transition frequencies (cyan diamonds), half-band transitions (black stars), both band and half-band transitions (blue squares), frequencies with non-negligible matrix element overlap (magenta triangles), and all frequencies in the band (red dots). (b) The best optimized shaking functions for the select frequency case corresponding to the magenta triangles in (a) for $n = 1$ (black, solid), $n = 2$ (blue, dot-dashed), and $n = 3$ (red, dashed).}
\end{figure}

We ran five different classes of optimization simulations. One class included every frequency in the band from DC up to $r = 0 \rightarrow r' = 5'$, another included only the $10$ band transition frequencies in this region, and a third included the $10$ half-band transition frequencies. All frequencies used here are tabulated in table \ref{tab:bandtrans}. A fourth simulation class included all $20$ band and half-band transition frequencies, and a fifth chose only the $9$ frequencies corresponding to appreciable ($>0.1$) matrix element overlap in figure \ref{fig:mat_el_total} (marked with asterisks in Table \ref{tab:bandtrans}). We plot the lowest error achieved after $1000$ iterations in figure \ref{fig:GAfreqs_pct_err}. Note that convergence below $10^{-3}\%$ is limited by phase errors in the split step method.

\begin{figure}[ht!]
\includegraphics[scale = .595]{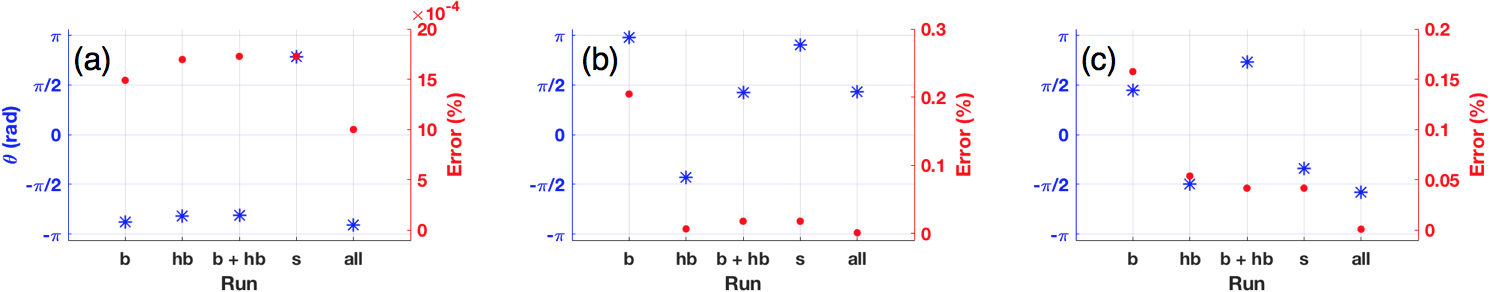}
\caption{\label{fig:GAfreqs_phase} Phase $\theta$ between the split arms (blue asterisks, left axis) and error (red dots, right axis) for (a) $n=1$, (b) $n=2$, and (c) $n = 3$. The best of ten runs for each of the five simulation classes is shown. The classes are labeled on the x-axis as follows: band transition frequencies (b), half-band transitions (hb), both band and half-band transitions (b+hb), select frequencies with non-negligible matrix element overlap (s), and all frequencies in the band (all)}
\end{figure}

From the results presented in figure \ref{fig:GAfreqs_phase}, we see that in all cases we can split the atom wavefunction to better than $0.3\%$. We see in figure \ref{fig:GAfreqs_pct_err}(a) that the error is lowest if we include all frequencies, but in this regime the interaction-induced heating (which is not present in our current simulation model) will be highest. For simulations restricted to the select strongest band transitions we can obtain splitting to better than $0.05\%$. While the use of more complex fitness functions (e.g. that used for first-order splitting in \cite{Anderson2017}) may further improve this splitting efficiency, we obtain good results by simply minimizing the error in equation \eref{eq:err}. In summary, by truncating our search space from $\approx 50$ frequencies (limited by the discrete temporal sampling inherent in the numerics) to $\approx 10$ frequencies, we can still split with sufficiently low error.

Even though we cannot access the relative phase $\theta$ of the two split arms of these optimized split states experimentally, it is of interest to examine them in simulation in order to better understand the shaking dynamics. Thus, we plot the final phase $\theta$ of the optimized split state for the best results of each of the five simulation classes and three splitting orders in figure \ref{fig:GAfreqs_phase}. We include the error from figure \ref{fig:GAfreqs_pct_err} for easy reference and comparison.

The results show that the phase dynamics of multi-frequency splitting are more complex than the simple model presented in section \ref{sec:pm_analytics} predicts. For example, when we split using the half-band transition frequencies we would expect that the phase $\theta$ be near zero. However, we find that this phase is closer to $\theta = -\pi$ for first-order splitting and $-\pi/2$ for higher-order splitting. Therefore, our simple model derived in the case of single-frequency shaking has broken down. Unfortunately it is analytically difficult to consider multiple frequency shaking due to the fact that neither the Jacobi-Anger expansion nor the results of Floquet analysis applies. Furthermore, as shown in figure \ref{fig:GAfreqs_pct_err}(b) the shaking function amplitude is about an order of magnitude greater than that used to justify the truncation of the sum in equation \eref{eq:V_PM_JA} and apply first-order perturbation theory.

We can, however, make some general inferences from our simple model. The fact that $|\theta| \approx \pi$ for the first-order split state likely comes as a result of the fact that the two-photon matrix element $|M^\mathrm{(c)}_{0,2}|^2$ is about a factor of two lower than the single-photon element $|M^\mathrm{(s)}_{0,1}|^2$ connecting $r = 0$ and $r' = 1$. Thus, transitions between the odd-parity states are more favorable, as can be seen from figure \ref{fig:shake_omega} where the dynamics of shaking at $\omega_{01}$ are far less complex than those of shaking at $\omega_{02}/2$.

For higher-lying states, the multiple pathways for an atom to get from $|r = 0\rangle$ to the higher-lying states means that the even and odd parity transitions tend to interfere, and the split state will in general be a mixture of the two bands. From figure \ref{fig:bands} we see that these higher-lying states $|r' > 2\rangle$ corresponding to splitting with $n>1$ are nearly degenerate. This interference causes the phase difference between the two split arms to average to $|\theta| = \pi/2$. In these cases the optimized split state is not dominated by population transfer into a single higher-lying band but rather this state is composed of a mixture of states corresponding to two nearly degenerate bands.

\begin{figure}[ht!]
\includegraphics[scale = .59]{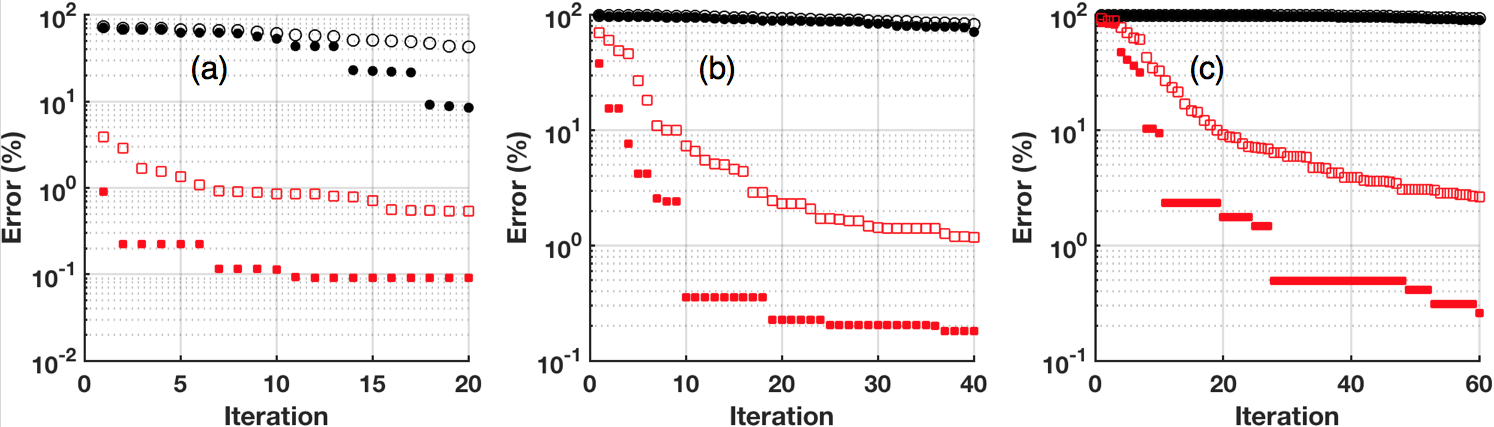}
\caption{\label{fig:learning_speed} Results of genetic optimization showing the mean (open markers) and best (closed markers) error for splitting with (a) $n = 1$, (b) $n = 2$ and (c) $n = 3$ versus number of iterations of the optimization algorithm. The black points indicate optimizations with all frequencies in the band from DC to $121$~kHz, and the red points indicate optimizations using only the truncated frequency space corresponding to the select band-to-band transitions indicated in table \ref{tab:bandtrans}. The error improvement is much faster with the truncated subspace.}
\end{figure}

From a purely experimental standpoint our results dramatically simplify the optimization landscape that we must explore. This allows for sufficient error minimization within a low number of iterations. That is, even though simulations with more frequencies tend to converge to lower errors, they take longer to do so. For example, if we run $10$ simulations to optimize splitting of the atom wavefunction with the select transition frequencies indicated in table \ref{tab:bandtrans}, for $n = 1, 2,$ and $3$, we can achieve convergence to better than $1\%$ error in (on average) $1$, $10$, and $28$ iterations, respectively. In each case, the error for the simulations with all frequencies in the band is $>70\%$, as shown in figure \ref{fig:learning_speed}. Figure \ref{fig:learning_speed} also shows that if we start with atoms in the $n = 2$ split state, we can optimize transfer into the $n = 3$ split state within $1\%$ within $<10$ iterations. In this case the total splitting time will double, but optimization of $6\hbar k_\mathrm{L}$ splitting is possible with fewer than $20$ total iterations. This learning speedup is extremely important for practical implementations of shaken lattice interferometry in that optimization happens more quickly and effectively, allowing for fast optimization of the interferometer sequence.

In the experiment, if the lattice depth is known (e.g. via the measurement scheme described in section \ref{sec:pm_analytics} or in \cite{Phillips2002, Raithel2009}) we can restrict our shaking the selected transitions with appreciable transition matrix elements. Then a closed-loop algorithm (e.g. the CRAB or dCRAB algorithm \cite{Montangero2011, Montangero2011b, Montangero2015}) can be used to efficiently optimize the splitting protocol (as was done for first-order splitting in \cite{Anderson2017b}). Thus we have found a reduced subspace that allows for faster optimization of the system and reduces heating due to off-resonant shaking \cite{Bloch2017}.

\section{\label{sec:conc} Conclusion}

In conclusion we demonstrate a simple means of using the band-to-band transitions to implement an atom beamsplitter in an optical lattice. We develop a theoretical model that allows us to use a single shaking frequency to implement low-order splitting. However, at higher frequencies our simple model breaks down and we must incorporate multiple frequencies in order to obtain good splitting. Due to heating caused by atom-atom interactions it is of interest to restrict our shaking frequencies to those resonant with single- and two-photon transitions between bands. We show that we can obtain higher-order splitting up to $n = 3$ with an error $<0.1\%$ by optimizing shaking with a learning algorithm using a reduced subspace of frequencies corresponding to the strongest band and half-band transition resonances. This simplification of the optimization landscape allows for faster optimization with less deleterious heating effects due to atom-atom interactions. Finally, we show that higher-order splitting can be implemented by accelerating the atoms in the optical lattice and can potentially allow for interferometry with sensitivity that scales with the cube of the interrogation time. This opens up potential new pathways for improving and expanding interferometry using atoms trapped in a shaken optical lattice.

\ack{The authors acknowledge funding from the NSF PFC under Grant No. 1125844 and the Northrop-Grumman Corporation and would like to thank D. Gu{\'e}ry-Odelin for fruitful discussions.}
\bigskip

\bibliography{full_text_incl_figures}

\end{document}